\documentstyle[12pt,epsfig]{article}

\input{tcilatex}

\begin{document}

\begin{center}
{\bf {\large Josephson versus Kondo coupling in a Quantum Dot Connected to
Two Superconductors}}

\vspace*{0.5cm} {\sl Gabriele Campagnano$^{\dagger }$, Domenico Giuliano$%
^{\dagger \dagger }$, Adele Naddeo{\large \footnote{{\sl {\large
{\footnotesize Corresponding author. Tel.: +39/081/676467}  }}
\par
{\sl {\large {\footnotesize E-mail address: adele.naddeo@na.infn.it}}}}}$%
^{\ddagger \ast }$, Arturo Tagliacozzo$^{\ddagger \ast }$}

\bigskip

{\small $^{\dagger }$ Department of NanoScience, Delft University of
Technology, Delft, The Netherlands}

{\small $^{\dagger \dagger }$ Dipartimento di Fisica, Universit\`{a} della
Calabria, Arcavacata di Rende - Cosenza, Italy}

{\small $~^{\ddagger }$ Istituto Nazionale di Fisica della Materia (INFM),
Unit\`{a} di Napoli, Napoli, Italy\\[0pt]
$^{\ast }$ Dipartimento di Scienze Fisiche, Universit\`{a} di Napoli
''Federico II'',\\[0pt]
Monte S.Angelo - via Cintia, I-80126 Napoli, Italy }
\end{center}

\vspace*{1cm}

\begin{center}
{\LARGE \ }

\vspace{8mm}{\bf Abstract}
\end{center}

\begin{quotation}
We apply a Gutzwiller-like variational technique to study Josephson
conduction across a quantum dot with an odd number of electrons connected to
two superconducting leads. Our method projects out all states on the dot but
the Kondo singlet and is valid when Kondo correlations are dominant and no
Andreev bound states localized at the dot are available for Kondo screening.
In these conditions superconducting pairing is a competing effect and the
junction is $\pi -$like, to optimize antiferromagnetic correlations on the
dot. As the superconducting gap increases, the Josephson current also
increases, but its phase dependence becomes strongly non sinusoidal.

\vspace*{0.5cm}


{\footnotesize PACS numbers: 85.25.-j, 72.10.Fk, 73.63.Kv\newpage } %
\baselineskip=18pt \setcounter{page}{2}
\end{quotation}

Phase coherent transport in hybrid superconducting-semiconducting
nanostructures is already extensively investigated. In these devices quite
interesting and surprising features emerge, due to electron-electron
interaction \cite{raimondi}. In this paper we study a Quantum Dot (QD) at
Coulomb Blockade (CB) with an odd number of electrons $N$, connected to
superconducting contacts \cite{proceedings}. Although such a kind of system
has not been realized yet, it is very likely that continuous improvement of
nanofabrication techniques will soon make it available. In particular, we
address the question whether the formation of a Kondo singlet at the dot
competes or cooperates with BCS s-wave pairing in the leads. We set up a
variational non perturbative approach which can be adopted as long as $%
\Delta <k_{B}T_{K}$, where $T_{K}$ is the Kondo temperature of the dot and $%
\Delta $ is the superconducting gap. We show how the flow to the Kondo fixed
point is affected by superconductivity because singlet pairing starts to
compete with Kondo correlations. The system behaves as a $\pi $-Josephson
junction \cite{spivak}, with a rather small critical Josephson current $I_{J}
$. As $\Delta $ increases and becomes comparable with $k_{B}T_{K}$, there is
a crossover to a regime in which Kondo correlations on the dot are very much
affected. The junction is still $\pi $-like and the Josephson critical
current increases but it becomes strongly non sinusoidal.

In our approach bound states localized within the dot area are responsible
for the bare magnetic moment on the dot. They are only weakly modified when
the contacts become superconducting, as they turn into Andreev bound states
within the gap. As discussed in the conclusions, the fact that our model
does not include other empty Andreev states hinders the possibility of
screening out the dot spin completely, in the large Kondo limit.

We believe that this is the reason why the $\pi $-coupling always wins in
our model, even when the ratio $\Delta /k_{B}T_{K}$ is quite small. Indeed $%
\pi $ coupling optimizes pair tunneling in the presence of strong
antiferromagnetic correlations on the dot.

Coulomb Blockade has been widely investigated in Quantum Dots with normal
leads \cite{kouwenhoven}. If the coupling with the contacts is weak and the
charging energy is higher than the thermal activation energy, DC conduction
is strongly dependent on the gate voltage $V_{g}$. The DC conductance across
the dot shows a sequence of peaks, occurring when $V_{g}$ provides the
energy to add an extra electron to the dot, taken from the contacts at
chemical potential $\mu $. As the dot is tuned in a ``Coulomb Blockade''
valley between two consecutive peaks, $N$ is fixed and the conductance is
heavily suppressed.

Kondo conductance may be achieved in a QD with $N$ odd within a CB valley by
increasing the coupling between dot and leads, provided the temperature $T$ $%
<T_{K}$ \cite{goldhaber}. At $T<T_{K}$, a strongly correlated state between
dot and leads sets in and a resonance in the density of states of the QD
opens up at $\mu $. Correspondingly, DC conductance across the dot
increases, until it eventually reaches the unitarity limit at $T=0$ \cite
{nozieres,silvano}.

Since charge dynamics is ``frozen out'' by CB, the QD in this regime is
usually modelized as a magnetic impurity with total spin $S=\frac{1}{2}$. A
Schrieffer-Wolff transformation on the single level Anderson impurity model
involving virtual zero or double occupancy of the impurity level leads to
the effective Kondo Hamiltonian with antiferromagnetic coupling between the
delocalized electrons of the leads and the impurity spin.

Magnetic impurities embedded in a bulk superconductor are known to strongly
influence the superconducting critical temperature. Adding impurity states
at energies below the superconducting gap can even give raise to gapless
superconductivity \cite{abrikosov}.

One can extend all this to the case of superconducting contacts. The
Schrieffer-Wolff transformation can be performed in the same way, provided $%
D>>\Delta $, where $D$ is the bandwidth of the itinerant electrons. However,
the resulting model is not equivalent to a single magnetic impurity in a
phase coherent superconductor. This is a crucial difference with respect to
the case of normal contacts when the Kondo screening cloud shows that the
phase coherence is established throughout the system.

On the contrary, in such a case the contacts retain their individual
superconducting properties and their individual phase for the order
parameter with a phase difference $\varphi $. The link between the two
superconductors is offered by the dot and is tunnel-like. This gives rise to
two important features:

1) Kondo phase coherence and superconducting phase coherence compete;

2) the system has the properties of a Josephson link between two different
superconductors.

The reason is that strong on site Coulomb repulsion makes the phase breaking
time induced by spin flip scattering processes much shorter than the Kondo
resonance lifetime $\hbar /k_{B}T_{K}$ \cite{proceedings}.

\begin{figure}[tbp]
\centering\includegraphics*[width=0.8\linewidth]{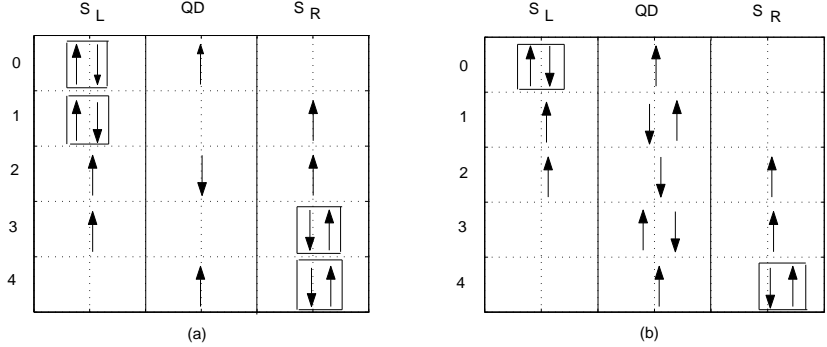}
\caption{Sketch of cotunneling processes ((a) hole process, (b) particle
process) in the perturbative regime, to move a Cooper pair from $L$ to $R$
across the QD. The sequence of steps $1\rightarrow 2\rightarrow 3$ is
constrained by Pauli restrictions. Analogous processes not depicted here can
take place, in which the spins are reversed.}
\label{figura1}
\end{figure}

Indeed, tunneling across a magnetic impurity between superconductors with on
site Coulomb repulsion has been recently revisited \cite
{matveev,spivak,clerk,vinokur}. If $\Delta \geq k_{B}T_{K}$, the system is
unable to scale toward the strongly-coupled Kondo regime. Sub-gap Cooper
pair tunneling is strongly suppressed by Coulomb repulsion, unless each
cotunneling step is accompanied by a spin flip at the impurity. Pair
tunneling can take place but it occurs in a three step sequence via virtual
empty or doubly occupied intermediate dot states, because of Pauli exclusion
principle, as depicted in Fig. \ref{figura1}. It has been proposed that such
a mechanism may reverse the sign of the Josephson current through the dot so
that the Josephson energy is at a minimum for $\varphi =\pi $, where $%
\varphi $ is the phase difference between the order parameters of the two
superconductors attached to the impurity ($\pi $-junction) \cite{spivak}.
This is indeed what we find with our variational approach.

Fig. \ref{figura1} mimicks Cooper pair tunneling via virtual breaking of a
pair, one at a time, with one quasiparticle moving into and out of the dot.
Particle and hole processes are allowed across the dot, with a mechanism
similar to the one involved in the Schrieffer-Wolff transformation. Here we
describe the hole process only; the particle one is fully analogous. Let us
consider an $\uparrow $ electron localized at the dot. In order to move a
pair from $L$ to $R$, we must first empty the dot. The sequence $%
1\rightarrow 2\rightarrow 3$ corresponds to an operator arrangement, $%
c_{L\uparrow }c_{R\downarrow }^{\dagger }c_{L\downarrow }c_{R\uparrow
}^{\dagger }$, which has a minus sign with respect to the standard
arrangement for Cooper pair tunneling: $c_{R\downarrow }^{\dagger
}c_{R\uparrow }^{\dagger }c_{L\uparrow }c_{L\downarrow }$ \cite{spivak}.
This produces a phase difference of $\pi $ with respect to the usual
Josephson coupling, which makes the system a ''$\pi -$junction''.

In Superconducting/Normal/Superconducting (SNS) structures, Cooper pair
subgap tunneling current takes place via Andreev states localized in the
normal region \cite{andreev}. When the normal region is given by a QD,
charge quantization at the dot is not spoiled as the contacts become
superconducting \cite{matgla}. The $I-V$ curve of a dot at CB between two
superconductors has been derived in \cite{zaikin}, showing an interplay
between multiple Andreev reflection and electron- electron interaction at
the QD.

We expect that fine tuning of the parameters of a S-QD-S device in an
orthogonal magnetic field may allow for the investigation of the full range
of physical conditions, from $\Delta \ll k_{B}T_{K}$ when the system can
flow toward the strongly coupled Kondo regime prior to the onset of
superconductivity in the leads, to $\Delta \gg k_{B}T_{K}$ when perturbation
theory holds \cite{vinokur}. In fact, the ground state degeneracy required
for Kondo coupling to take place can be obtained by driving the dot to a
level crossing by means of an applied magnetic field \cite
{silvano,noi,nazarov,nos}.

We study the $T=0$ case with $\Delta <k_{B}T_{K}$ using a non perturbative
variational technique. The fact that a global phase coherence cannot take
place in the system justifies our variational approach outlined in the
following. Hence, we add Kondo correlations to a state which is the
superposition of two BCS states for the left and the right contact with
different phases of the order parameter. Such a technique has already been
applied to the case of a dot with normal contacts, and it has been shown to
provide good qualitative results in the perturbative regime as well, where
it reproduces the poor man's scaling equation \cite{nos}.

To construct the trial state, we start from the state $|\varphi ,s\rangle $,
given by the product of the left ($L$) condensate times the right ($R$)
condensate, with the two order parameters having a phase difference $\varphi
$, times the state of the dot $|s\rangle $:
\begin{equation}
|\varphi ,s\rangle =|{\rm BCS},\;L\rangle {{\times} }|{\rm BCS}%
,\;R,\;\varphi \rangle {{\times} }|s\rangle .
\end{equation}
\noindent Here $s$ is the spin component along the quantization axis of the
dot spin ${\bf S}_{d}=\frac{1}{2}$ at CB with odd $N$.

The minimal model for the Kondo interaction between electrons localized on
the QD and electrons from the contacts is $H_{K}=J\vec{\sigma}(0){{\cdot} }%
\vec{S}_{d}$. The spin density operator of the delocalized electrons, $\vec{%
\sigma}(0)$, at the position of the dot $x=0$ along the vertical axis, is:
\[
\vec{\sigma}(0)=\frac{1}{2{\cal {V}}}\sum_{q,q^{^{\prime }}}(c_{L,q,\sigma
}^{\dagger }+c_{R,q,\sigma }^{\dagger })\vec{\tau}_{\sigma ,\sigma
^{^{\prime }}}(c_{L,q^{^{\prime }},\sigma ^{^{\prime }}}+c_{R,q^{^{\prime
}},\sigma ^{^{\prime }}}).
\]
\noindent We have used the fermion operators $c_{q,j}(c_{q,j}^{\dagger })$, $%
(j=L,R)$ in the plane wave representation to describe the contacts particles
and ${\cal {V}}$ is the normalization volume of the leads. We take the
symmetric case in which hybridization of the dot with the $L$ and $R$
contacts, $\Gamma $, is the same. Kondo coupling is antiferromagnetic (AF): $%
J>0$.

The total Hamiltonian is $H=H_{S}+H_{K}$, where $H_{S}$, defined in eq. (\ref
{bcsh}) below, is the Hamiltonian for the $L$ and $R$ superconducting
contacts in the $BCS$ approximation.

The correlated trial state is constructed by applying a Gutzwiller-like
projector $P_{g}$ to $|\varphi ,s\rangle $. As $T\rightarrow 0$, the system
scales towards the strongly coupled, large $J$ regime. Correspondingly, $%
P_{g}$ gradually projects out the high-energy components of the trial state,
so that eventually only a localized spin singlet survives at the QD.

The ``projector'' $P_{g}$ is defined as \cite{nos}:
\[
P_{g}=\left( 1-\frac{3}{4}g\right) +g(\vec{\sigma}(0))^{2}-4{\bf S}_{d}{{%
{\cdot}} }\vec{\sigma}(0),
\]
\noindent where $g$ is a variational parameter which ranges between $g=0$
and $g=4/3$. When $g=0$ we have $P_{0}=1-4{\bf S}_{d}{{\cdot} }\vec{\sigma}%
(0)$; $P_{0}$ fully projects out the high energy localized spin triplet at $%
x=0$. As $g$ varies from $0$ to $4/3$ also the localized spin doublet state
it increasingly projected out. Eventually, when $g$ reaches the value $g=4/3$%
, only the localized spin singlet is left over.

The trial state is defined as:

\begin{equation}
|g,\varphi \rangle =P_{g}|\varphi ,s\rangle .  \label{eq2}
\end{equation}
\noindent The value of $g(J)$ is determined by finding the minimum of the
energy functional $E[g,J,\varphi ,\Delta ]$, defined as:

\begin{equation}
E[g,J,\varphi ,\Delta ]=\frac{\langle g,\varphi |H|g,\varphi \rangle }{%
\langle g,\varphi |g,\varphi \rangle }.  \label{eq4}
\end{equation}
\noindent Eq. (\ref{eq4}) can be calculated by first expressing the products
$P_{g}HP_{g}$ and $(P_{g})^{2}$ in terms of the usual fermion quasiparticle
operators $\alpha _{j,q}(\alpha _{j,q}^{\dagger }),\beta _{j,q}(\beta
_{j,q}^{\dagger })$, $(j=L,R)$ which destroy (create) excitations on the BCS
states of the $L$ and $R$ contact and then by normal-ordering the
corresponding operator products. The operators are \cite{gabri}:
\begin{eqnarray}
\alpha _{j,q} &=&u_{q}c_{j,q,\uparrow }-v_{q}e^{i\phi
_{j}}c_{j,-q,\downarrow }^{\dagger }  \nonumber \\
\beta _{j,-q} &=&u_{q}c_{j,-q,\downarrow }+v_{q}e^{i\phi
_{j}}c_{j,q,\uparrow }^{\dagger },  \label{eq1}
\end{eqnarray}
where $u_{q}$ and $v_{q}$ are the BCS coherence factors and $\phi
_{R}=\varphi $ while $\phi _{L}=0$. The Hamiltonian for the contacts $H_{S}$
is conveniently expressed in terms of these operators as:
\begin{equation}
H_{S}=E_{{\rm BCS}}+\sum_{q,j=L,R}E_{q}(\alpha _{j,q}^{\dagger }\alpha
_{j,q}+\beta _{j,-q}^{\dagger }\beta _{j,-q}).  \label{bcsh}
\end{equation}
Here $E_{{\rm BCS}}$ is the total ground state energy of the condensates and
$E_{q}=\sqrt{\xi _{q}^{2}+\Delta ^{2}}$ are the energies of the
quasiparticle excitations with $\xi _{q}=q^{2}/2m-\mu $ (we put $\hbar
=k_{B}=1$ throughout the paper).

The variation of the energy \cite{gabri} due to the AF coupling at the
quantum dot in units of the bandwidth $D$, $\epsilon \lbrack \xi ,j,\delta
,\varphi ]\equiv E[g,J,\Delta ,\varphi ]/D$, takes a simple form once
expressed in terms of the parameters $\xi =\frac{1}{2}\left( 1-\frac{3}{4}%
g\right) $, $j=3J/4D$ and $\delta =\Delta /D$:
\[
\epsilon \lbrack \xi ,j,\delta ,\varphi ]=-\frac{j}{2}\frac{1-\frac{1}{2}%
\delta ^{2}(1+\cos (\varphi ))/N^{2}(0)\lambda ^{2}}{(1+\xi ^{2})+(\xi
^{2}-1)\delta ^{2}(1+\cos (\varphi ))/(2\left( N(0)\lambda \right) ^{2})}+
\]
\begin{equation}
+\frac{(1-\xi +\xi ^{2})\left[ \sqrt{1+\delta ^{2}}-\delta ^{2}\cos (\varphi
)/N(0)\lambda \right] }{(1+\xi ^{2})+(\xi ^{2}-1)\delta ^{2}(1+\cos (\varphi
))/(2\left( N(0)\lambda \right) ^{2})}.  \label{eq5}
\end{equation}
\noindent

Here $\lambda $ is the BCS electron-electron interaction strength and $N(0)$
is the normal phase density of states at the Fermi level for each spin
polarization. We will take $N(0)\lambda =0.3$ throughout the paper. The
first term is the expectation value of $H_{K}$, while the second is the
raise in the average value of the kinetic energy Hamiltonian, $H_{S}$, due
to the formation of the singlet between the QD and the contacts. The value
of $\xi _{{\rm min}}$, at which the energy is at a minimum, measures how
much higher-energy spin states are projected out: $\xi _{{\rm min}}=0$
corresponds to full projection of states different from a localized spin
singlet at the impurity.

At $\delta =0$ the strong coupling fixed point is $j\rightarrow \infty $ and
$\xi _{{\rm min}}(j)\rightarrow 0$. In the flow to the fixed point, there is
a large decrease of the Kondo energy:
\begin{equation}
\Omega _{K}\equiv \epsilon [ \xi ,j,0 , 0 ] =\frac{1-\xi _{{\rm min}}+(\xi _{%
{\rm min}})^{2}-\frac{j}{2}}{ 1+(\xi _{{\rm min}})^{2}}.  \label{omk}
\end{equation}

\begin{figure}[tbp]
\centering \includegraphics*[width=0.7\linewidth]{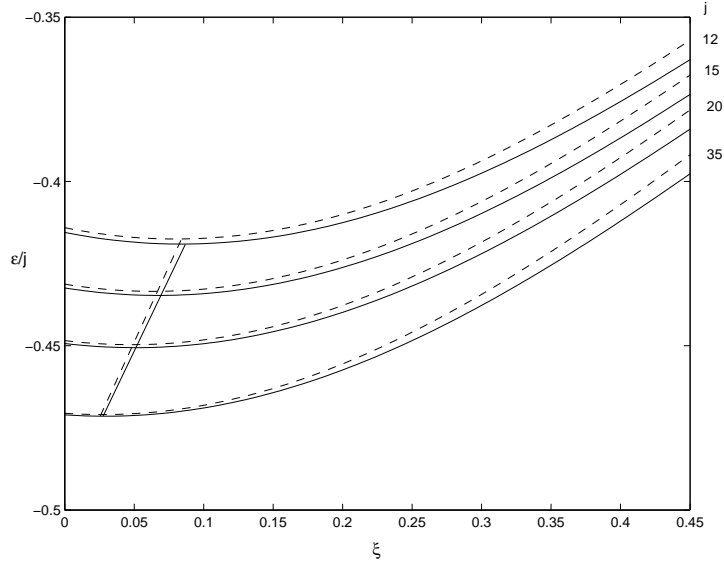}
\caption{$\protect\epsilon /j$ vs. $\protect\xi $ for different $j$ and $%
\protect\delta =0.06$ in the two cases $\protect\varphi =0$ (broken line)
and $\protect\varphi =\protect\pi $ (full line). For large Kondo coupling $j$
the minimum of the curve moves toward $\protect\xi =0$. }
\label{figura2}
\end{figure}

At $\delta \neq 0$, we follow the minimum of the energy of the correlated
state, $\epsilon ^{\ast }[j,\delta ,\varphi ]\equiv \epsilon \lbrack \xi _{%
{\rm min}},j,\delta ,\varphi ]$, in the flow $j\rightarrow \infty $. In Fig.
\ref{figura2} we plot $\epsilon /j$ vs. $\xi $ for different $j$ at $\delta
=0.06$. We see that the minimum of $\epsilon \lbrack \xi ,j,\delta ,\varphi ]
$ is located at smaller and smaller $\xi $ values the larger the Kondo
correlations are ($j\rightarrow \infty $). However the zero value for $\xi _{%
{\rm min}}$, corresponding to the full projection of states different from a
localized spin singlet on the dot, is never reached. Energies corresponding
to $\varphi =\pi $ (full line) are always lower than those corresponding to $%
\varphi =0$ (broken line). This is a general feature of our results: the $%
\pi $-junction behavior is estabilished throughout the whole range of $%
\delta $ values, until $\Delta $ reaches the Kondo scale $T_{K}$.

\begin{figure}[tbp]
\centering \includegraphics*[width=0.7\linewidth]{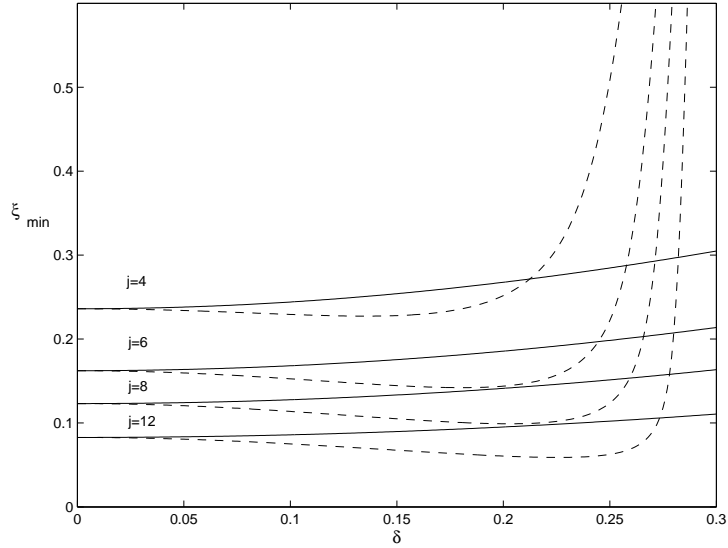}
\caption{$\protect\xi _{{\rm min}}$ vs. $\protect\delta $ for various $j$,
for $\protect\varphi =0$ (broken line) and $\protect\varphi =\protect\pi $
(full line). In the case $\protect\varphi =0$, for small values of $\protect%
\delta $, $\protect\xi _{{\rm min}}$ is lowered, showing that a small amount
of superconductivity singlet pairing favors Kondo correlations. On the other
hand, for large $\protect\delta $, $\protect\xi $ increases abruptly, what
signals the disruption of the Kondo correlations. By contrast, the weakening
of the Kondo correlations is much steadier but less sharp when $\protect%
\varphi =\protect\pi $.}
\label{figura3}
\end{figure}

\begin{figure}[tbp]
\centering \includegraphics*[width=0.9\linewidth]{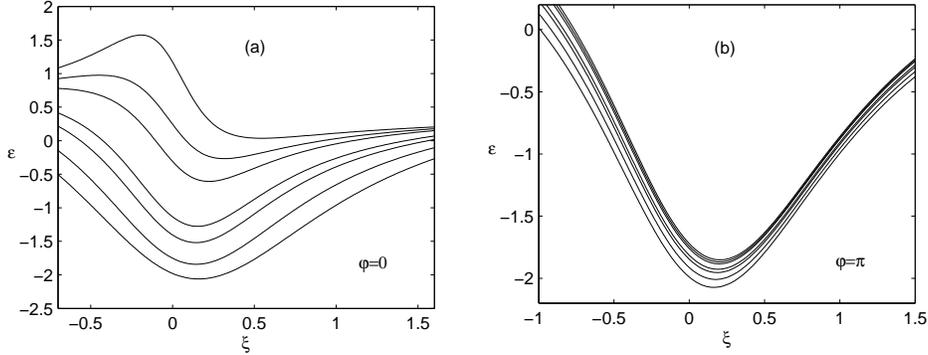}
\caption{$\protect\epsilon $ vs. $\protect\xi $ for $\protect\delta %
=0.05,0.15,0.20,0.22,0.25,0.26,0.27$ in the two cases $\protect\varphi =0$
(a) and $\protect\varphi =\protect\pi $ (b). (a) Deeper minima correspond to
smaller $\protect\delta $. For $\protect\delta $ close to $0.3$ the system
will not be able to build up the Kondo singlet. (b) Deeper minima correspond
to smaller $\protect\delta $.}
\label{figura5}
\end{figure}

In Fig. \ref{figura3} we plot $\xi _{{\rm min}}$ vs. $\delta $ for $\varphi
=0,\pi $ and various $j$ values. In the case $\varphi =0$, $\xi _{{\rm min}}$
grows up abruptly for $\delta >0.25$ what shows that the choice $\varphi =0$
strongly disfavours the Kondo correlations, as soon as the superconducting
pairing correlation length $\xi _{S}$ decreases. The dependence of $\xi _{%
{\rm min}}$ on $\delta $ is much weaker when $\varphi =\pi $. Inspection of
Fig. \ref{figura5} shows that two regimes can be envisaged as $\delta $
increases:

\vspace*{0.5cm}

{\sl $a)$ $\Delta \ll T_{K}$: Superconducting pairing in the leads is a
perturbation on the Kondo correlated state }

\begin{figure}[tbp]
\centering \includegraphics*[width=0.9\linewidth]{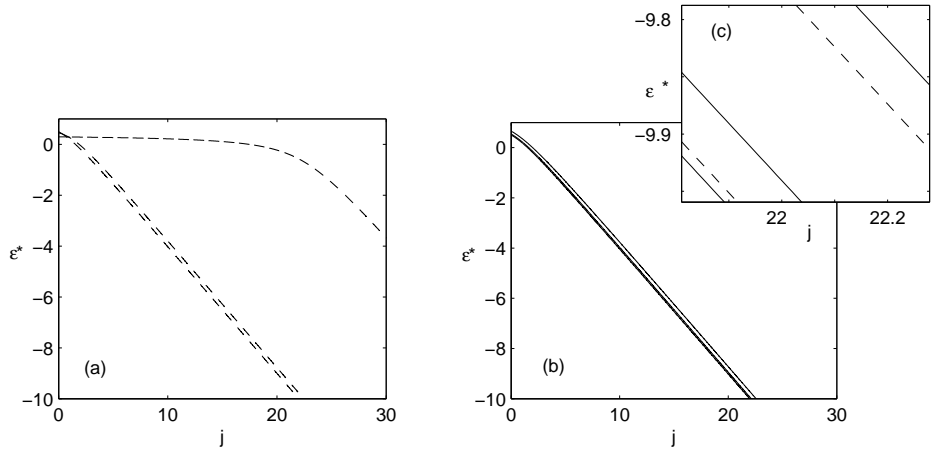}
\caption{$\protect\epsilon ^{\ast }$ vs. $\jmath $ for $\protect\delta %
=0.05,0.15,0.29$ in the two single cases $\protect\varphi =0$ (a) and $%
\protect\varphi =\protect\pi $ (b) and together (c) for a comparison.}
\label{figura4}
\end{figure}

For small values of $\delta $ the minimum of the energy can be attained both
for $\varphi =0$ and for $\varphi =\pi $. According to Fig. \ref{figura3},
there is a slight decrease of $\xi _{{\rm min}}$ when $\delta $ increases
for $\varphi =0$. Hence, the superconducting singlet correlations seem to
cooperate for small $\delta $'s in such a case. By contrast, $\xi _{{\rm min}%
}$ increases steadly with $\delta $ for $\varphi =\pi $, what shows that
singlet correlations on the impurity (with characteristic correlation length
$\xi _{K}$) compete with singlet pairing but are much less disruptive.
This is confirmed by Figs. \ref{figura4}a), \ref{figura4}b). The energy
minimum is strongly affected when $\delta $ increases for $\varphi =0$,
while this is not the case for $\varphi =\pi $. Again, the case $\varphi
=\pi $ is always favoured in energy for any $j$ value (see Fig. \ref{figura4}%
c)).

The energy of the correlated state $\epsilon ^{\ast }[j,\delta ,\varphi ]$,
to second order in $\delta $, is:
\begin{eqnarray}
\epsilon ^{\ast }[j,\delta ,\varphi ] &=&\Omega _{K}\left\{ 1+\delta ^{2}
\left[ \frac{1}{2}+\frac{1-(\xi _{{\rm min}})^{2}}{(N(0)\lambda )^{2}}-\frac{%
\xi _{{\rm min}}^{2}}{(N(0)\lambda )^{2}}\cos \varphi \right] \right\} +
\nonumber \\
&&+\frac{j}{2}\frac{\delta ^{2}}{1+(\xi _{{\rm min}})^{2}}\left[ \frac{1}{2}+%
\frac{1}{(N(0)\lambda )^{2}}\right] .  \label{eq7}
\end{eqnarray}
\noindent Because $\Omega _{K}<0$ for large $j$, it is clear that $\varphi
=\pi $ is favoured.

\vspace*{0.5cm}

{\sl $b)$ $\Delta \leq T_{K}$: Superconductivity strongly competes with
Kondo ordering }

Such a situation is better shown in Fig. \ref{figura5}a): when the phase
difference between the two superconducting leads is $\varphi =0$, the energy
functional, as $\delta $ approaches the limit value $0.3$, cannot gain a
minimum. Indeed the system cannot develop a Kondo singlet on the impurity
site. Conversely, Fig. \ref{figura5}b) shows that a minimum always exists
when the phase difference is $\varphi =\pi $.

It is easy to derive the Josephson current, according to:

\[
I_{J}=2e\partial \epsilon ^{\ast }[j,\delta ,\varphi ]/\partial \varphi .
\]
\noindent

\begin{figure}[tbp]
\centering \includegraphics*[width=0.9\linewidth]{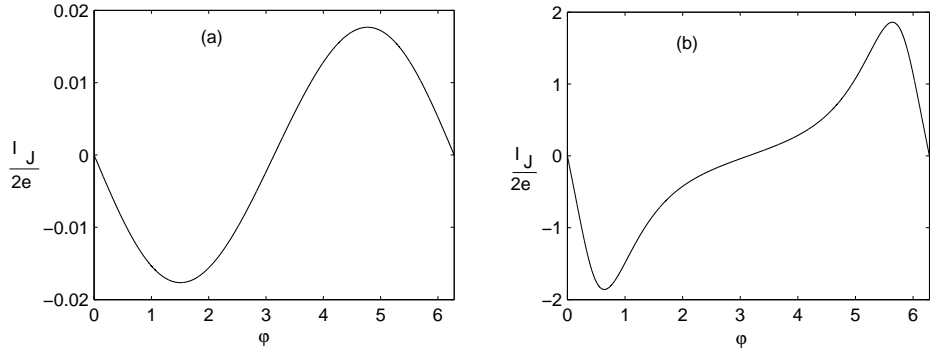}
\caption{Different possible regimes for the Josephson current at zero
temperature for $j=8$: a) $\Delta \ll T_{K}$; b) $\Delta \leq T_{K}$.}
\label{figura6}
\end{figure}

In regime $a)$, $I_{J}$ it is still sinusoidal, of the form $I_{J}=2e\frac{%
\Omega _{K}\delta ^{2}\xi _{{\rm min}}^{2}}{(N(0)\lambda )^{2}}\sin \varphi $%
. The sign of such a current is reversed with respect to the conventional
one for Josephson systems, that is, the device behaves as a $\pi $-junction.
In such a regime superconductivity starts to compete with Kondo correlations
and the Josephson critical current comes out to be rather small, as we can
see in Fig. \ref{figura6}a).

$\pi $-coupling wins in our model no matter how small the ratio $\Delta
/k_{B}T_{K}$ is. This is counterintuitive because one expects that, when
Kondo correlations are fully estabilished and the magnetic moment at the dot
is totally screened, the scattering of Cooper pairs should be potential-like
only. This implies that no phase breaking takes place at the tunneling and
the junction should behave as a conventional Josephson junction, with its
energy minimum at $\varphi =0$ \cite{clerk}. This is not the case here
because the density of states does not include extra subgap Andreev states
localized at the dot which could provide the screening of the magnetic
moment. We expect that, if Andreev bound states are accounted for properly
in the model, they can provide the required full Kondo screening for
vanishing $\Delta /k_{B}T_{K}$ ratio.

As $\delta $ increases, we enter a new regime in which superconductivity
strongly competes with Kondo correlations. The Josephson critical current
becomes now sizeable. When $\delta $ exceeds a threshold value (close to $0.3
$) the Josephson current develops strong non sinusoidal components (see Fig.
\ref{figura6}b)).

The regime $\Delta >T_{K}$, in which the superconducting order is dominant,
cannot be descibed by our variational Ansatz. In this case, the Kondo
interaction is better treated perturbatively. This regime was studied in
ref. \cite{clerk} using a $NCA$ perturbative approach and in ref. \cite
{vinokur}. Their results can be interpreted in terms of Andreev bound states
in the normal region.

In the noninteracting limit the single particle energy spectrum for the
S-N-S sandwich has two Andreev levels corresponding to states localized in
the normal region with energy within the gap. Taking the Fermi level $\mu $
as the reference energy, the one of the two states with positive (negative)
energy is mostly particle (hole)-like. In this case, Josephson conduction
across the normal region involves both Andreev states. Because the levels
are non degenerate Kondo correlations cannot take place until $J$ is large
enough. Kondo correlations require that the particle and hole-like states
cross each other, so that screening of the impurity spin can take place at
the dot site.

In conclusion, we have generalized the variational approach introduced in
ref. \cite{nos} to study Josephson conduction through a quantum dot at
Coulomb blockade connected to two superconducting leads. Within such an
approach, we analyzed the region of parameters where the dot lays in the
strongly coupled Kondo regime and showed that, in this regime, it behaves as
a $\pi $-junction with a small Josephson current. Our formalism shows that
the onset of $\pi $-junction regime takes place much before
antiferromagnetic correlations at the dot can be treated perturbatively,
that is, much before Kondo effect has been disrupted by superconductivity.

\pagebreak


\begin{thebibliography}{99}
\bibitem{raimondi}  C. J. Lambert and R. Raimondi, {\it J. of Phys. Condens.
Matter }{\bf 10}, 901 (1998).

\bibitem{proceedings}  G. Campagnano, D. Giuliano, A. Naddeo, A. Tagliacozzo
- ''{\it Josephson versus Kondo Coupling at a Quantum Dot with
Superconducting Contacts}'' - Proceedings of EUCAS 2003-6th European
Conference on Applied Superconductivity, Sorrento (NA), Italy, 14-18
September 2003, Institute of Physics (IOP) Publishing, Bristol, UK (2003),
to appear.

\bibitem{spivak}  B. I. Spivak and S. A. Kivelson, {\it Phys. Rev. B} {\bf %
43 }, 3740 (1991).

\bibitem{kouwenhoven}  L. P.~Kouwenhoven {\sl et al.}, in ``{\it Mesoscopic
electron transport}'', L.~Sohn, L. P.~Kouwenhoven and G.~Sch\"{o}n eds.,
NATO ASI Series {\bf E 345},105, Kluwer, Dordrecht, Netherlands (1997).

\bibitem{goldhaber}  D.~Goldhaber-Gordon {\it et al.}, {\it Nature} {\bf 391}
, 156 (1998); S. M.~Cronenwett, T. H.~Oosterkamp and L. P.~Kouwenhoven, {\it %
Science} {\bf 281}, 540 (1998).

\bibitem{nozieres}  P. Nozi\'{e}res, {\it J. Low Temp. Phys.} {\bf 17}, 31
(1974).

\bibitem{silvano}  S.~Sasaki, S.~De Franceschi, J. M.~Elzerman, W. G.~van
der Wiel, M.~Eto, S.~Tarucha and L. P.~Kouwenhoven, {\it Nature} {\bf 405},
764 (2000).

\bibitem{abrikosov}  A. A. Abrikosov and L. P. Gorkov, {\it Zh. Eksp. Teor.
Fiz.} {\bf 39}, 1781 (1960).

\bibitem{matveev}  L. I. Glazman and K. A. Matveev, {\it Pis'ma Zh. Eksp.
Teor. Fiz.} {\bf 49} (10), 570 (1989).

\bibitem{clerk}  A. A. Clerk and V. Ambegaokar, {\it Phys.Rev.B} {\bf 61},
9109 (2000).

\bibitem{vinokur}  V. I. Kozub, A. V. Lopatin and V. M. Vinokur, {\it Phys.
Rev. Lett.} {\bf 90}, 226805 (2003).

\bibitem{andreev}  A. F. Andreev, {\it Zh. Eksp. Teor. Fiz.} {\bf 46}, 1823
(1964), {\bf 22}, 655 (1965).

\bibitem{matgla}  K. A. Matveev and L. I. Glazman, {\it Phys. Rev. Lett.}
{\bf 81}, 3739 (1998).

\bibitem{zaikin}  Y. Avishai, A. Golub and A. D. Zaikin, {\it Phys. Rev. B}
{\bf 63}, 134515 (2001).

\bibitem{noi}  D. Giuliano and A. Tagliacozzo, {\it Phys. Rev. Lett.} {\bf %
84 }, 4677 (2000).

\bibitem{nazarov}  M. Eto and Y. Nazarov, {\it Phys. Rev. Lett.} {\bf 85},
1306 (2000).

\bibitem{nos}  D. Giuliano, B. Jouault and A. Tagliacozzo, {\it Phys. Rev. B}
{\bf 63}, 125318 (2001).

\bibitem{gabri}  G. Campagnano, Tesi di laurea in Fisica, {\it Universita'
di Napoli ''Federico II''}, (2001) unpublished.
\end{thebibliography}
\end{document}